# Monetizing Propaganda:
# How Far-right Extremists Earn Money by Video Streaming

Megan Squire
Department of Computer Science
Elon University, msquire@elon.edu

Video streaming platforms such as Youtube, Twitch, and DLive allow users to live-stream video content for viewers who can optionally express their appreciation through monetary donations. DLive is one of the smaller and lesser-known streaming platforms, and historically has had fewer content moderation practices. It has thus become a popular place for violent extremists and other clandestine groups to earn money and propagandize. What is the financial structure of the DLive streaming ecosystem and how much money is changing hands? In the past it has been difficult to understand how far-right extremists fundraise via podcasts and video streams because of the secretive nature of the activity and because of the difficulty of getting data from social media platforms. This paper describes a novel experiment to collect and analyze data from DLive's publicly available ledgers of transactions in order to understand the financial structure of the clandestine, extreme far-right video streaming community. The main findings of this paper are, first, that the majority of donors are using micropayments in varying frequencies, but a small handful of donors spend large amounts of money to finance their favorite streamers. Next, the timing of donations to high-profile far-right streamers follows a fairly predictable pattern that is closely tied to a broadcast schedule. Finally, the far-right video streaming financial landscape is divided into separate cliques which exhibit very little crossover in terms of sizable donations. This work will be important to technology companies, policymakers, and researchers who are trying to understand how niche social media services, including video platforms, are being exploited by extremists to propagandize and fundraise.

**ACM Reference Format:**

## 1   Introduction

DLive is a video streaming service similar to Twitch and Youtube that allows users ("streamers", "content creators") to record themselves talking, playing video games, and so on, and for other users to watch and interact with this content as it happens either through text chat or by offering a financial donation. Financial transactions between streamers and donors are managed on DLive via a site currency called *lemons*. Lemons inserted into and extracted from the DLive system are logged on a publicly available ledger of transactions.

Systematically studying the DLive platform in order to understand the video streaming space has several advantages. First, because DLive is a smaller and lesser-known streaming platform than its competitors such as Twitch and Youtube, and because its content moderation is far less strict than the larger and better-known platforms, DLive has become a popular place for violent extremists - and other groups which have been removed from more mainstream services - to earn money while producing propaganda. Second, DLive transaction data is publicly available. In the past it has been difficult to understand far-right financing via podcasts and video streams both because of its nature as a clandestine activity, and because of the difficulty of getting data from the mainstream social media platforms. Because DLive provides a publicly available ledger of transactions, it represents a unique opportunity to collect data about the secretive world of extremist financing via video streams.

The purpose of this paper is to describe the collection and analysis of data from the DLive system in order to answer two research questions about the far-right extremist video streaming ecosystem, as follows:

- RQ1: Timing and Size. When do the financial transactions happen, and how much money is changing hands? Are there interesting patterns in the timing or quantity of donations, or cash-in and cash-out behaviors?
- RQ2: Network. What does the network of payers and payees look like? What are the key structural features of the payment network?

The remainder of this paper shows how we answered these questions, first by reviewing the background and prior work that inspired this research, then by writing software to collect publicly available data from the DLive system, and finally by analyzing that data. Before summarizing our findings, we review the limitations of this approach and suggest some avenues for future work.

Because this work is the first of its kind to provide systematic data collection around the financial aspects of the right-wing extremist video streaming network, we expect that our results will be useful to social media researchers, platform providers, and technology policymakers, many of whom have raised important questions about how the monetization of extremist propaganda takes place.

## 2   Background and related work

This work was inspired by literature about far-right extremist financing in an online context, as well as some prior research about the celebrity culture and fundraising opportunities presented by video livestreaming platforms.

The use of donations as a funding vehicle for far-right extremist groups and actors is not a new phenomenon and is not limited to the online context [1]. However, the emergence of newer online "content subscription and livestreaming platforms" has improved access to these services, and with it the funding potential for hate groups. Prior work such as [2] describes the fundraising potential presented by donation-only platforms like Patreon and Subscribestar and content delivery services like DLive, and concludes that 14 groups were using these services to fundraise, with three hate groups specifically using the DLive service.

We were also interested in how influencer culture and celebrity culture on video streaming sites such as Youtube [3] has been adopted by far-right extremists [4, 5], increasingly to monetize their extreme political views [6]. The donation infrastructure adopted by DLive was closely modeled after Youtube's "Superchat" feature [7] in which users can use tips or donations to elevate their chat messages and gain attention from video streamers. The enduring question as yet unanswered by the literature is about the donation potential in these online fundraising spaces, so this is the main purpose of our study. Some prior work [8] has shown that donation behavior is closely correlated with viewing behavior, which is also something we will investigate further in Section 4.3.

## 3   Data collection

The data for this project was gathered from the publicly available DLive transaction ledger via its web-based application programming interface (API) located at https://graphigo.prd.dlive.tv. Our data collection software connected to the API and retrieved data for each of 119 DLive user accounts, divided into three main categories:

1. Streamers. The far-right extremist streamer category consists of 55 individuals or groups who regularly create English-language, far-right extremist content and receive donations on the DLive platform. Included in this group are individuals who have been profiled by journalists and watchdog groups as espousing far right extremist ideas, and who have been banned from other video streaming platforms for promoting hatred and violence. Examples include Nick Fuentes [9], Vincent James [10], Patrick Casey [11], Martin Sellner [12], and so on.
2. Mega-Donors. This data set consists of 20 DLive users who have donated in large amounts to the far-right streamers, including "tbased", "gio949", and "doomersquidward". These donors in some cases reach their own levels of notoriety within the streaming community, for example they may be granted chat moderator privileges, or they may have high numbers of followers despite not producing their own content.
3. People of Interest (POI). We collected data for 44 people classified as far-right extremists who may not have started streaming or donating on DLive but nonetheless have an account there and have dabbled on the site by creating a profile page which has attracted followers. Many of the POI are well-known personalities or organizations on the far right, for example Counter Currents Publishing [13], VDARE [14], and so on. Like people in the 'streamer' category, these individuals are already well-known for promoting hatred or racially and ethnically motivated violence on other social media and streaming platforms. The biggest difference between the 'streamer' category and the 'POI' category is whether the individual has actually begun streaming on the DLive site or whether they have just created an account for later use.

## 3.1 Data format

For each of these 119 accounts, we retrieved the ledger of earnings, as well as information about their follower/following relationships. Each of these API results were saved as a file, then the results were parsed, and the data was uploaded to a relational database for further querying and analysis. Data has been extracted from this system and is available at [29].

Below is an example of a transaction record from the ledger of earnings. Each time the ledger is downloaded for a user account, the entire transaction history is retrieved, back to the launch date of the current DLive blockchain on April 16, 2020. The *createdAt* is a Unix epoch timestamp. The *amount* is given in lemon points, which must be divided by 100,000 to arrive at the actual number of lemons transacted. Note that *amount* reflects a net value, after the DLive fee of 25% has been removed from the donation, as shown in the example below:

{"seq": "18142931", "txType": "Donation In", "createdAt": "1610071357000", "description": "@patrickcasey donated 1000 LEMON(s) to post dlive-93233454+FoAMbx-Mg", "amount": "75000000", "balance": "221666649947", "__typename": "Transaction"}

## 3.2 Summary Statistics

Table 1 summarizes key features of this data set.

**Table 1:** Key features of data set

| Data set feature | |
|---|---|
| First timestamp represented in data | 2020-04-16 01:21:02 GMT |
| Last timestamp represented in data | 2021-02-06 19:50:11 GMT |
| Count of user accounts | 119 |
| Count of collection activities | 13 |
| First date of collection | 2020-06-30 |
| Final data of collection | 2021-02-06 |

Table 2 shows the different financial transaction types represented in the data, and the count of transactions of that type. The final two categories, *Subscription Income* and *Subscription Paid* appear to have been introduced in October 2020.

Table 2: **Transaction types represented in data set**

| Transaction type | Count of transactions |
|---|---|
| Cash In | 28200 |
| Cash Out | 3261 |
| Donation In | 965344 |
| Donation Out | 49448 |
| Subscription Income | 2497 |
| Subscription Paid | 3 |

Table 3 compares each of the three user types (streamer, mega-donor, and person of interest) in terms of their relative numbers of outbound transactions (donations *to* someone), inbound transactions (donations *from* someone), and number of followers. Since streamers are actively using the platform to create content, they tend to have high numbers of inbound donations and high follower counts. Mega-donors, on the other hand, tend to have high numbers of outbound transactions, as donations, but fewer inbound transactions. Users designated as "people of interest" (POI) are those who have created an account on the system, but have not begun streaming there very much, and thus have low transaction counts. However, as they are often high-profile e-celebrities on other streaming platforms, POI often have high follower counts.

Table 3: **Characteristics of each user type on DLive**

| User type | # Txns: Outbound | # Txns: Inbound | # Followers |
|---|---|---|---|
| Streamer | Lower | Higher | Higher |
| Donor | Higher | Lower | Lower |
| Person of Interest | Lower | Lower | Higher |

# 4 Data Analysis

Our first research question (RQ1) was about the timing and size of transactions. We use the DLive transaction types (Cash-In, Cash-Out, Donation-In, and Donation-Out) to organize these findings in sections 4.1-4.3. Section 4.4 discusses RQ2 and the network of donors and streamers.

## 4.1 Cash-In Transactions

There are several categories of transactions designated as "Cash-In" in the DLive ledger. Table 4 shows the relative popularity of each of these Cash-In types. Lemon points are converted to lemons by dividing by 100,000, then into USD using the $0.012 conversion rate given at [15].

Table 4: **Cash-In subtypes on DLive**

| Cash-In type | # Txns | Sum of Lemon Points | Lemons to USD |
|---|---|---|---|
| Referral bonus | 20018 | 1324443000 | $158.93 |
| Cash In (purchasing lemons) | 4631 | 2385175400000 | 286,221.05 |
| Watch/chat reward | 2884 | 6316764803 | 758.01 |
| Subscription cashback | 360 | 2427800000 | 291.34 |
| [blank description] | 141 | 54947549797 | 6593.71 |
| Chest refund | 81 | 1620500000 | 194.46 |
| Ads rewards | 51 | 510000 | 0.06 |
| Refund | 34 | 108903974947 | 13068.48 |

Because currency ("lemon") purchases represent the largest category of Cash-In in our data set, we provide additional analysis of these transactions. Lemons are sold in bundles either in the Web browser or via the mobile app, and the bundles cost different amounts depending on which payment method is used. Table 5 shows a comparison of lemon costs. The cost to purchase a lemon on the app store is around 1.7 cents, whereas on the browser it is between 1.2 and 1.7 cents, depending on whether the user pays with Amazon Pay or cryptocurrency.

**Table 5:** Comparison of currency costs on DLive (Cash-In)

| Mobile Device Cash-In | | Browser-based Cash-In | | |
|---|---|---|---|---|
| # lemons in bundle | Cost, Apple app store (unit price) | # lemons in bundle | Cost, Amazon Pay (unit price) | Cost, Cryptocurrency (unit price) |
| 57 | $0.99 (0.017368) | 88 | $1.53 (0.017386) | $1.06 (0.012045) |
| 174 | 2.99 (0.017184) | 288 | 4.04 (0.014028) | 3.46 (0.012014) |
| 582 | 9.99 (0.017165) | 688 | 9.06 (0.013169) | 8.26 (0.012006) |
| 874 | 14.99 (0.017151) | 1188 | 15.34 (0.012912) | 14.26 (0.012003) |
| 2041 | 34.99 (0.017144) | 2888 | 36.67 (0.012697) | 34.66 (0.012001) |
| 5832 | 99.99 (0.017145) | 7888 | 99.44 (0.012606) | 94.66 (0.012001) |
| | | 78888 | | 946.66 (0.012000) |

Table 6 shows the number of users from each category purchasing each bundle size. Unsurprisingly, users classified as mega-donors purchase more lemons and purchase larger bundle sizes than the users classified as streamers or POI. Only 4 users purchased the largest 78888-lemon bundle available only using cryptocurrency.

**Table 6:** Comparison of bundle sizes purchased by each category of user

| Mobile Device Cash-In | | | | Browser-based Cash-In | | | |
|---|---|---|---|---|---|---|---|
| # lemons in bundle | Streamers | Donors | POI | # lemons in bundle | Streamers | Donors | POI |
| 57 | 0 | 21 | 7 | 88 | 0 | 21 | 1 |
| 174 | 6 | 199 | 15 | 288 | 1 | 117 | 8 |
| 582 | 2 | 181 | 23 | 688 | 0 | 147 | 1 |
| 874 | 0 | 85 | 6 | 1188 | 1 | 126 | 8 |
| 2041 | 3 | 377 | 11 | 2888 | 3 | 190 | 8 |
| 5832 | 3 | 910 | 0 | 7888 | 1 | 2063 | 1 |
| | | | | 78888 | 0 | 4 | 0 |

Table 7 shows the top 15 users by Cash-In transactions to purchase lemons. All of these users were classified as mega-donors in our user classification system. The total lemon purchase in USD and the percentages of browser-based versus mobile device is given in the table.

**Table 7:** Top 15 Users, by sum of Cash-In (purchasing lemons) transaction value

| Username | # Cash-In Txns | Sum of Lemon Points | Lemons to USD | % browser | % mobile |
|---|---|---|---|---|---|
| tbased | 993 | 751388400000 | $90,166.61 | 100 | 0 |
| doomersquidward | 697 | 375948900000 | 45,113.87 | 6 | 94 |
| Psylosopher | 293 | 226948400000 | 27,233.81 | 100 | 0 |
| gio949 | 674 | 177485000000 | 21,298.20 | 4 | 96 |
| based-dollar | 192 | 143179600000 | 17,181.55 | 100 | 0 |
| freedombob | 124 | 104911200000 | 12,589.34 | 100 | 0 |

| Username | # Cash-In Txns | Sum of Lemon Points | Lemons to USD | % browser | % mobile |
|---:|---:|---:|---:|---:|---:|
| pyrrhus777 | 119 | 91527200000 | 10,983.26 | 100 | 0 |
| finkelstein | 106 | 82701600000 | 9,924.19 | 99 | 1 |
| mcpaddy | 96 | 67544800000 | 8,105.38 | 100 | 0 |
| CascadiaPNW | 448 | 58297800000 | 6,995.74 | 12 | 88 |
| tc457823 | 122 | 50818300000 | 6,098.20 | 99 | 1 |
| OberGroyper | 72 | 47584000000 | 5,710.08 | 50 | 50 |
| Medraut | 112 | 42195600000 | 5,063.47 | 100 | 0 |
| AireBear3.0 | 40 | 30772000000 | 3,692.64 | 100 | 0 |
| groyptech | 40 | 29212000000 | 3,505.44 | 100 | 0 |

This data shows that most mega-donors purchase lemons from a single platform: either browser or mobile but not both. In addition, some mega-donors do *not* seem to make their purchase decisions based on finding the lowest unit price ("most lemons for the money"). Instead, some mega-donors are willing to pay a premium by choosing to purchase using the app store, and by repeatedly purchasing many small bundles rather than a few lower-priced large bundles. Future work could investigate whether this behavior is related to impulse purchasing or mobile device dependence or both.

There is very little information available about users on the DLive platform other than their usernames and whatever statements they choose to publish on their public profile pages, so this Cash-In data provides some of the only clues about the motivation and behavior of donors on the platform.

## 4.2 Cash-Out Transactions

Transactions designated as Cash-Out are divided four main types, as shown in Table 8.

**Table 8:** Cash-Out subtypes on DLive

| Cash-Out type | # Txns | Sum of Lemon Points | Lemons to USD |
|---:|---:|---:|---:|
| Add to chest | 2680 | 59104100000 | $7,092.49 |
| CashOut (extract cash) | 546 | 6990942100000 | 838,913.05 |
| User requested refund | 26 | 24741600000 | 2,968.99 |
| Account suspended | 9 | 275886139546 | 33,106.34 |

The "Add to chest" transaction type is the largest category in terms of the count of transactions, however the total in currency is relatively small, only $7,092.49. As a livestream progresses, a virtual treasure chest begins accumulating [16]. Streamers can also add lemons to the chest manually. The streamer can then open the treasure chest after at least 5 lemons are in it, and those lemons will be distributed to the viewers watching. Some streamers open the chest multiple times during a broadcast, others wait until the end, and others do not use this feature at all.

The "CashOut" category – largest in terms of dollar amounts - is users simply extracting money from the system. It stands to reason that users we classify as mega-donors and POI do not have notable CashOut activity, whereas the streamers do perform many CashOut activities. Table 9 shows the breakdown of those 546 CashOut (extract cash) transactions in terms of their user profile. The 55 streamers we tracked were able to extract over $800,000 from DLive during the 9-month period we followed these accounts.

**Table 9:** CashOut (extract cash) frequency by user type

| User type | # Txns | Sum of Lemon Points | Lemons to USD |
|---:|---:|---:|---:|
| Streamer | 526 | 6914913000000 | $829,789.56 |
| Donor | 0 | 0 | 0.00 |
| Person of Interest | 20 | 76029100000 | 9,123.49 |

Table 10 shows the streamers whose transactions indicate they have cashed out more than $10,000 from the system, as well as whether those streamers had any "account suspended" transactions. The nine "account suspended" transactions were all carried out on January 20, 2021. Five of these transactions were for the same streamer, nickjfuentes. The 26 "user requested refund" transactions were all carried out on February 2, 2021, and were all for the same donor, doomersquidward.

**Table 10: Top Users by CashOut (extract cash), and account suspended**

| Cash-Out (Extract Cash) | | | | Cash-Out (Account Suspended) | | |
|---|---|---|---|---|---|---|
| Username | # Txns | Sum of Lemon Points | Lemons to USD | # Txns | Sum of Lemon Points | Lemons to USD |
| nickjfuentes | 16 | 782476100000 | 93,897.13 | 5 | 250487824947 | 30,058.54 |
| PatrickCasey | 24 | 665527400000 | 79,863.29 | | | |
| OwenBenjaminComedy | 31 | 664262300000 | 79,711.48 | | | |
| PeteSantilli | 24 | 597262300000 | 71,671.48 | | | |
| JadenMcNeil | 21 | 551951200000 | 66,234.14 | | | |
| theralphretort | 23 | 359925900000 | 43,191.11 | | | |
| SubCultured | 14 | 276258900000 | 33,151.07 | | | |
| BeardsonBeardly | 28 | 248041800000 | 29,765.02 | | | |
| shalit | 13 | 229785100000 | 27,574.21 | | | |
| RedIceTV | 11 | 228410000000 | 27,409.20 | | | |
| tedwang | 7 | 224886000000 | 26,986.32 | | | |
| VincentJames | 6 | 221219400000 | 26,546.33 | 1 | 18994711100 | 2,279.37 |
| JesseLeePeterson | 20 | 158408100000 | 19,008.97 | | | |
| BakedAlaska | 14 | 154827700000 | 18,579.32 | 1 | 206900000 | 24.83 |
| REALPOSEIDON | 25 | 144078600000 | 17,289.43 | | | |
| LiftTheVeil | 11 | 124290000000 | 14,914.80 | | | |
| VPFM | 10 | 115580200000 | 13,869.62 | | | |
| Franssen | 23 | 108020800000 | 12,962.50 | | | |
| WhiteRabbitRadio | 20 | 106410600000 | 12,769.27 | | | |
| MartinSellner | 12 | 95647500000 | 11,477.70 | | | |
| Azzmador | 29 | 92763200000 | 11,131.58 | | | |
| Mr_White_Tuber | 18 | 90593400000 | 10,871.21 | | | |
| MurderTheMedia | 7 | 86190700000 | 10,342.88 | 1 | 3454184500 | 414.50 |

## 4.3  Donation-In, Donation-Out Transactions

Whereas Cash-In and Cash-Out transactions represent money flowing into and out of the DLive system, Donations-In and Donations-Out represent currency being transferred between users within the system. We calculate a grand total of Additionally, Donations-In and Donations-Out are approximately the inverse of one another, minus a transaction cost taken by DLive itself. For example, in the transaction ledger, a 100-lemon donation from User A to User B will appear as a 100-lemon Donation-Out from User A and a 75-lemon Donation-In to User B. DLive has changed its policies over time on the amount it reserves from each donation and what that amount is used for [17], but the end result is that the donations are logged in both directions, minus a transaction fee, and donations could be reflected twice in the transaction ledger if both sender and recipient are in our collection of 119 user accounts.

We find that among the users in this data set, the users with the highest counts for "Donation-In" are largely the same as the biggest "Cash-Out" users (shown in Table 10). Similarly, the highest counts for "Donation-Out" are – for the most part – the same as the biggest Cash-In users (shown in Table 7).

Do people donate one lemon at a time, or do they donate in bundles? The DLive interface offers the option to donate in bundles, and there is also the option to pay for a monthly channel subscription which comes with extra perks. Table 11 shows the most frequently used bundles for the users in this data set, both inbound and outbound.

**Table 11: Commonly purchased donation bundles**

| Bundle Size, in lemons | Name of Bundle | Qty purchased as "donation in" | Qty purchased as "donation out" |
|---|---|---|---|
| 1 | Lemon | 349527 | 5432 |
| 10 | Ice Cream | 152025 | 3588 |
| 100 | Diamond | 181789 | 12574 |
| 298 | Subscription | 51487 | 10176 |
| 1000 | Ninjaghini | 27421 | 5506 |
| 10000 | Ninjet | 1653 | 948 |

Table 11 deepens the assertions about inbound and outbound donations introduced in Table 3. It is the streamers in our data set who are the recipients of inbound donations, and here we show that the majority of these inbound donations come in smaller bundles such as 1, 10, or 100 lemons. Conversely, the majority of the outbound donations in our data set come from users we classified as mega-donors, and we find the majority of their outbound donations are in the 100 and 298 lemon range.

In terms of timing of donations, we posit that donation activity will be closely tied to a user's streaming schedule. This is because donations are both an expression of general support for a streamer, and also a reflection of how users feel about the content of their livestream. Unfortunately, the DLive earnings API does not provide data about a user's streaming activity on the platform. Thus, to explore the relationship between the timing of donations and livestreaming activity, we turn to a streamer who has both a predictable schedule for which historical data exists, and who also has a large number of donations on the DLive platform: the user "nickjfuentes". Recall from Table 10 that this individual is the largest earner in our data set. His posted schedule for streaming is Monday through Friday evenings (Central Time), and the four-year historical record of his "America First" video streams is available at [18].

Figure 1 shows the daily donation amounts for nickjfuentes from April 15, 2020 to January 9, 2021 when his DLive account was suspended [19] for "inciting violent and illegal activities" on the platform. The timestamp of each transaction was converted from GMT to GMT-5, to approximate Central Daylight Time for the majority of transactions.

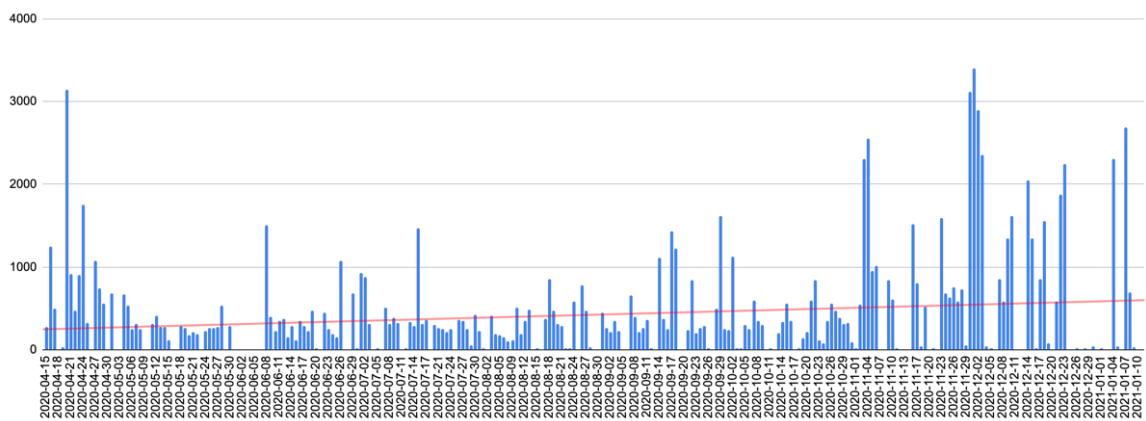

**Figure 1:** NickJFuentes donations per day (USD) April 2020 – January 2021, with trendline

For most of the year, his inbound donations line up with his weeknight streaming schedule, and weekends are easy to discern as gaps on the graph. Table 12 shows his inbound donations by day of week.

Table 12: **NickJFuentes inbound donations by day of week, April 2020 – January 2021**

| Day of Week (GMT-5) | # Txns | Sum of Lemon Points | Lemons to USD |
|---|---|---|---|
| Monday | 28754 | 239664303100 | $28,759.74 |
| Tuesday | 27051 | 191812103300 | 23,017.48 |
| Wednesday | 17367 | 171835538600 | 20,620.29 |
| Thursday | 22166 | 185594248500 | 22,271.32 |
| Friday | 19081 | 129733330300 | 15,568.02 |
| Saturday | 2347 | 17034855800 | 2,044.18 |
| Sunday | 1602 | 14129698400 | 1,695.57 |

The exceptions to this periodicity come at the end of the year when his schedule became irregular due to attending "in real life" (IRL) events for the "Stop the Steal" campaign [20] and holiday breaks. During his final week on the platform, the week of January 4, 2021, he also took a few days off, for example to attend the Capitol insurrection [21].

Despite taking numerous days off at the end of the year, the donation average for nickjfuentes actually went *up* following the election. Figure 1 shows the daily moving average increased from $272 in April to almost $600 by the end of the year. The highest donation totals for the nickjfuentes user coincided with his pro-Trump election coverage (November 3 and 4) and his promotion and coverage of the false claim that the election results were fraudulent. He began promoting his election fraud coverage using a series titled "Election War." For example, the show titled "ELECTION WAR: Election Fraud EXPOSED in Public Hearing" on December 1 earned him $3,396 in donations and the show titled "ELECTION WAR: Smoking Gun EXPOSED in Georgia" earned another $2,349 on December 3.

We posit that other streamers may show the same close relationship between their streaming activities and their donations, allowing us to reverse-engineer a streaming history from the donation timing in this data set, even if the recording of a stream were removed from the platform. One exception to this might be the 298-lemon inbound donations for monthly subscriptions since those could happen at any time of day and are not necessarily closely tied to streaming activity.

## 4.4 Network of donors and streamers

In this section we address RQ2, which covers the network of donors and streamers in the DLive system. The payment relationship, or donations between users, provides the basis for this network graph.

The nodes in the network are DLive users, and the edges are the donations made from one user to another. This network includes transactions of type "Donation In," so this means any DLive user who donated to one or more of the 119 core users is a potential node on the graph. The edges on the graph reflect the direction of the payments from one user to another. To keep the graph size manageable, only edges with donations worth more than 10000 lemons (about $120) are included. The final network graph is thus comprised of 1092 nodes and 1581 directed edges. Figure 2 shows the network as drawn in Gephi [22] with a Yifan Hu [23] network layout. Node size corresponds to degree (larger nodes = more connections) and colors correspond to user roles (streamers = pink and donors = gray).

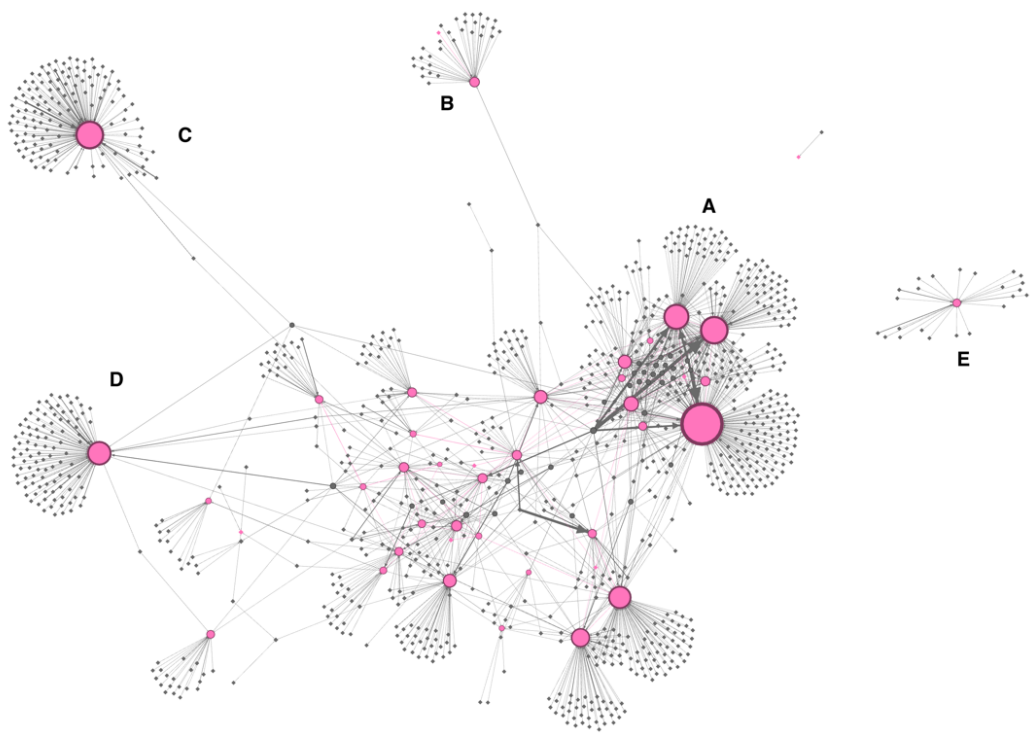

**Figure 2:** Network of donors (gray) and streamers (pink), node size = degree, minimum donation $120

There are three subgraphs in the network: two small, disconnected ones and then the main graph. One feature that stands out about this network is the large number of star-shaped structures: streamers and their donors are clearly identifiable as a pink circle with radiating spokes. Many of these stars are barely connected to the rest of the graph and may only have one or two donors tying them to the main network. We conclude that the far-right extremist streamer donation network on DLive is thus very fractured, at least at the $120 minimum donation level. Users who are prepared to give at least $120 tend to pick one, possibly two, streamers to whom to donate this much money.

We added labels to a few of the discrete fan communities in this network. The cluster marked A represents a large set of streamers and donors calling themselves "Groypers" and associated with streamers nickjfuentes, PatrickCasey, and JadenMcNeil (see Table 10) and mega-donors tbased and doomersquidward (see Table 7). The cluster marked B is comprised of the Proud Boys-affiliated streaming account called MurderTheMedia [24] and its donors. The C cluster represents anti-government activist and radio host PeteSantilli [25] and his donors. The D cluster is OwenBenjaminComedy and his donors, who call themselves "Bears" [26]. The E cluster, disconnected from the main graph, represents donors to MartinSellner, an Austrian with ties to the perpetrator of the 2019 Christchurch, New Zealand massacre [12]. His livestreams are typically in German which may explain the disconnection from the rest of the graph.

Figure 3 shows the graph colorized by sub-communities in the network, as determined using Gephi's modularity metric, with a default resolution of 1. The modularity algorithm finds 14 separate communities in the network. Interestingly, the Groyper community marked "A" in Figure 3 has been subdivided into two separate modules in this version of the diagram. One of these groups, shown in blue in Figure 3 corresponds to the nickjfuentes wing, with doomersquidward as a key mega-donor. The other group, shown in orange, corresponds to the mega-donor called tbased and his two funding recipients, PatrickCasey and JadenMcNeil. It came as no surprise that following the Capitol insurrection on January 6th when these groups and individuals were facing intense scrutiny for their planning and participation in those riots, the Groypers essentially turned on one another and began fracturing into two camps [27], similar to those indicated on this diagram.

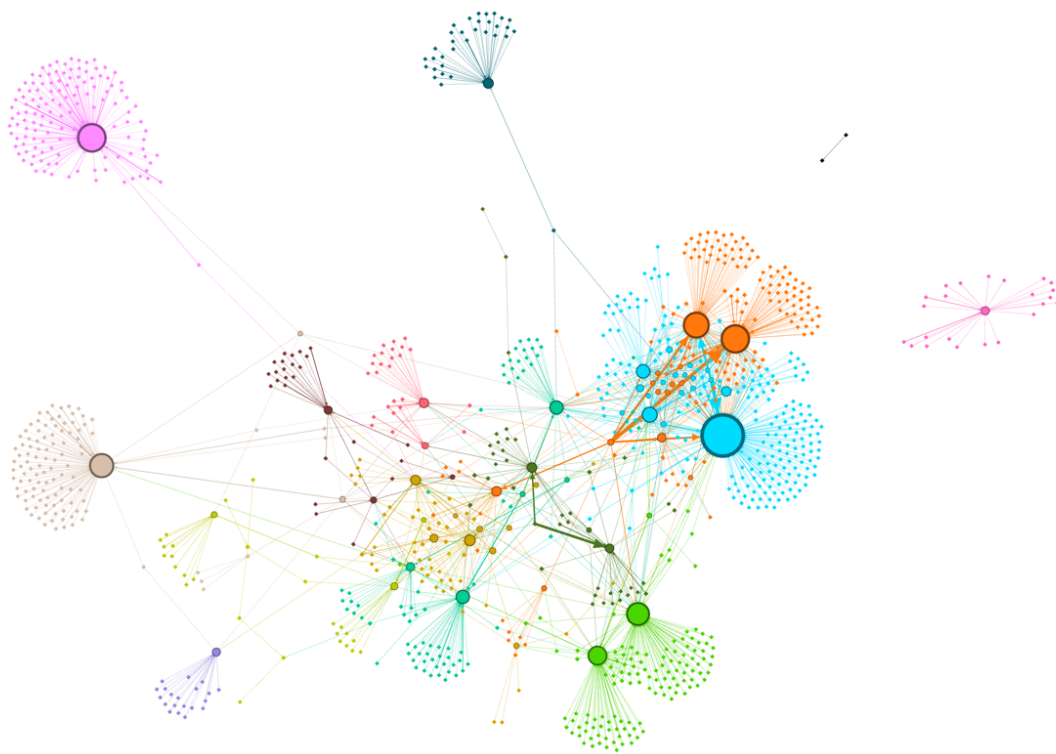

**Figure 3:** Network of donors and streamers, colorized by modularity (14 communities), node size = degree

# 5 Limitations and future work

This study has some limitations and opportunities for expansion in future work. First, we did not thoroughly analyze the followers/following relationships data that we collected from the DLive API, and we did not yet attempt to connect those relationships to financial transactions. The relationships data was the key characteristic of the users we classified as "people of interest" and these users were not well-represented in the financial transactions at all, so looking more closely at their relationships would help explain this category of users. We could also examine the relationship between donations and follower/following behavior, for example, to answer whether streamers bother to follow their mega-donors.

Second, because this data set is limited to 119 users, we are potentially missing some interesting figures in the network, particularly in the streamer category. DLive does not advertise that these people use their platform, so finding streamers was sometimes a challenge. That said, the group of far right extremist streamers is a minority on the DLive platform, and this data set is at least representative of that small population.

Similarly, we would have liked to have expanded our discussion of the links between donations and streaming schedule. We only used one streamer for this discussion in Section 4.4 (nickjfuentes). One challenge in confirming the relationship between donations and a streaming schedule is the lack of a historical record of precise streaming start and end times. Lacking these, we chose to focus on a streamer with an approximate schedule and a lot of data in the system.

In terms of future work, because DLive has made a number of changes to its platform in the wake of the January 6, 2021 insurrection at the US Capitol building, many of the questions we asked and answered here will not be able to be expanded. For example, DLive announced the de-monetization of political streaming channels [28] and they deleted the accounts of several of the 55 streamers in this data set who participated in or promoted the insurrection [19].

# 6   Conclusions

In this work we have attempted to shine a light on the clandestine world of far-right extremists earning money via video livestreams. Throughout 2020 and into 2021, these content creators used a platform called DLive to produce live video content and interact with their audiences via chat and taking donations, cementing the importance of "monetized propaganda" as a lucrative funding channel. The fact that DLive has a publicly viewable transaction ledger means that it is possible to construct a systematic study of this type of funding mechanism for the first time. We chose to focus on far-right extremist actors not only because they were very effective at using this novel streaming vehicle to earn money and they were ubiquitous on the DLive platform, but because of the potential danger presented by large amounts of cash being injected into their cause.

Findings from this study illuminate the size and scope of the issue – how much money is really changing hands in a donation-based livestreaming environment – and also the variety of far-right actors willing to exploit this environment. We show that with a regularly produced livestream show on a niche platform like DLive, far-right actors can earn over $100,000 in donations in less than a year. This money is available by courting both mega-donors and smaller donors. The funds are able to be cashed out regularly from the platform, providing a form of regular income to political extremists. Further, we show that multiple flavors of far-right actors – from the "Groypers" of America First to the Proud Boys to the "Bears" who follow Owen Benjamin – were able to financially exploit this platform, even working separately. It is our hope that by revealing the size and scope of what is possible for financial exploitation of niche platform like DLive, we will be able to inspire similar work on other platforms which may also have open data available.

## ACKNOWLEDGMENTS
We would like to thank Daniel Hosterman for providing invaluable assistance with understanding the DLive API and for providing code examples that showed how to connect to this API.

## REFERENCES
[1]	Tom Keatinge, Florence Keen & Kayla Izenman. 2019. Fundraising for Right-Wing Extremist Movements, The RUSI Journal, 164:2, 10-23, DOI: 10.1080/03071847.2019.1621479
[2]	Institute for Strategic Dialogue. 2020. Bankrolling Bigotry: An Overview of the Online Funding Strategies of American Hate Groups. October 27. available at https://www.isdglobal.org/wp-content/uploads/2020/10/bankrolling-bigotry-3.pdf
[3]	Alice E. Marwick. 2015. You May Know Me from YouTube: (Micro-)Celebrity in Social Media. In A Companion to Celebrity (eds P.D. Marshall and S. Redmond). https://doi.org/10.1002/9781118475089.ch18
[4]	Rebecca Lewis. 2020. "This is what the news Won't show you": YouTube creators and the reactionary politics of micro-celebrity. Television & New Media, 21(2), 201-217.
[5]	Rebecca Lewis. 2018. Alternative influence: Broadcasting the reactionary right on YouTube. Data & Society, 18.
[6]	Giovanni De Gregorio and Catalina Goanta. 2020. The Influencer Republic: Monetizing Political Speech on Social Media (November 4). Available at SSRN: https://ssrn.com/abstract=3725188 or http://dx.doi.org/10.2139/ssrn.3725188
[7]	Ishmael Daro and Craig Silverman. 2018. How YouTube's "Super Chat" System Is Pushing Video Creators Toward More Extreme Content. Buzzfeed. May 17. https://www.buzzfeed.com/ishmaeldaro/youtube-comments-hate-speech-racist-white-nationalists-super
[8]	Donghee Yvette Wohn and Guo Zhang Freeman. 2020. Live streaming, playing, and money spending behaviors in eSports. Games and Culture, 15(1), 73-88.
[9]	Ben Lorber. 2021. "America First Is Inevitable" Nick Fuentes, the Groyper Army, and the Mainstreaming of White Nationalism. Political Research Associates. January 15. Available at https://www.politicalresearch.org/2021/01/15/america-first-inevitable
[10]	Frank John Tristan. 2017. How Lake Forest's Alt-Right Media Collective the Red Elephants Twists Truth to Make Fake News. OC Weekly. August 31. Available at https://ocweekly.com/how-lake-forests-alt-right-media-collective-the-red-elephants-twists-truth-to-make-fake-news-8382380/
[11]	Joan Braune. 2019. Void and Idol: A Critical Theory Analysis of the Neo-fascist>Alt-Right. Journal of Hate Studies, Vol. 15 (No. 1, 2019), pp. 11–37.
[12]	Heidi Beirich, and Wendy Via. 2020. "Generation Identity". Global Project against Hate and Extremism. https://www.politico.eu/wp-content/uploads/2020/07/GPAHE-July-2020-Report.pdf
[13]	Southern Poverty Law Center. n.d. Extremist profile: Greg Johnson. https://www.splcenter.org/fighting-hate/extremist-files/individual/greg-johnson

[14]	Jeff Tischauser and Kevin Musgrave. 2020. Far-Right Media as Imitated Counterpublicity: A Discourse Analysis on Racial Meaning and Identity on Vdare.com. Howard Journal of Communications, 31(3), 282-296.
[15]	DLive. n.d. Welcome Letter. https://community.dlive.tv/about/welcome-letter
[16]	DLive, n.d. Treasure Chest. https://help.dlive.tv/hc/en-us/articles/360039325311-Treasure-Chest
[17]	DLive. 2020. Change in BTT Staking Rewards Distribution. December 15. https://community.dlive.tv/2020/12/15/change-in-btt-staking-rewards-distribution/
[18]	The TV Database. Available at: https://www.thetvdb.com/series/america-first-with-nicholas-j-fuentes/allseasons/official
[19]	DLive. 2021. Building a safe and welcoming community. January 9. https://community.dlive.tv/2021/01/09/building-a-safe-and-welcoming-community/
[20]	Malachi Barrett. 2021. "Far-right activist who encouraged U.S. Capitol occupation also organized 'stop the steal' rally in Michigan" MLive. January 7. https://www.mlive.com/politics/2021/01/far-right-activist-who-encouraged-us-capitol-occupation-also-organized-stop-the-steal-rally-in-michigan.html
[21]	Dan Frosch, Rachael Levy and Zusha Elinson. 2021. "Extremists in Capitol Riot Had Histories of Violent Rhetoric and Threats," Wall Street Journal. January 14. https://www.wsj.com/articles/extremists-in-capitol-riots-had-histories-of-violent-rhetoric-and-threats-11610639781
[22]	Mathieu Bastian, Sebastien Heymann and Mathieu Jacomy. 2009. Gephi: an open source software for exploring and manipulating networks. In Proceedings of the International AAAI Conference on Web and Social Media (Vol. 3, No. 1).
[23]	Yifan Hu. 2005. Efficient, high-quality force-directed graph drawing. Mathematica Journal, 10(1), 37-71.
[24]	Department of Justice. 2021. Texas Man, Self-Proclaimed Leader of Honolulu Proud Boys Now Indicted by a Federal Grand Jury for Conspiracy to Obstruct Congress. February 3. Available at: https://www.justice.gov/usao-dc/pr/texas-man-self-proclaimed-leader-honolulu-proud-boys-now-indicted-federal-grand-jury
[25]	Conrad Wilson. 2016. Pete Santilli – Journalist Or Felon? Oregon Public Broadcasting. March 2. Available at https://www.opb.org/news/series/burns-oregon-standoff-bundy-militia-news-updates/pete-santilli-charges-legal-case-free-speech/
[26]	Claire Goforth. 2021. Banned 'alt-right' comedian returns to spread coronavirus misinformation. Daily Dot. January 27. Available at https://www.dailydot.com/debug/owen-benjamin-youtube-twitter/
[27]	Will Sommer and Kelly Weill. 2021. FBI Informant Panic Is Ruining Friendships All Over the Far Right. The Daily Beast. https://www.thedailybeast.com/fbi-informant-panic-is-ruining-friendships-all-over-the-far-right-proud-boys-and-america-firsters
[28]	DLive. 2021. An Open Letter to the DLive Community. January 17. https://community.dlive.tv/2021/01/17/an-open-letter-to-the-dlive-community
[29]	Megan Squire. 2021. DLive project data. https://github.com/megansquire/dliveProjectData